\documentclass{Interspeech}

\interspeechcameraready

\title{Speaker Targeting via Self-Speaker Adaptation for Multi-talker ASR}
\author[affiliation={1}]{Weiqing}{Wang}
\author[affiliation={1}]{Taejin}{Park}
\author[affiliation={1}]{Ivan}{Medennikov}
\author[affiliation={1}]{Jinhan}{Wang}
\author[affiliation={1}]{Kunal}{Dhawan}
\author[affiliation={1}]{He}{Huang}
\author[affiliation={1}]{Nithin}{Rao Koluguri}
\author[affiliation={1}]{Jagadeesh}{Balam}
\author[affiliation={1}]{Boris}{Ginsburg}

\affiliation{}{NVIDIA}{USA}
\email{\{weiqingw,taejinp,imedennikov,jinhanw,kdhawan,heh,nkoluguri,jbalam,bginsburg\}@nvidia.com}
\keywords{Multi-talker ASR, Multi-speaker ASR, Target-speaker ASR, Streaming ASR
}

\usepackage{comment}

\usepackage{multirow}      
\usepackage{booktabs}      
\usepackage{amssymb}       
\usepackage{pifont}        
\usepackage{array}         
\usepackage{caption}       
\usepackage{makecell}
\usepackage{marvosym}
\usepackage{url}

\begin{document}

\maketitle

\begin{abstract}
We propose a self-speaker adaptation method for streaming multi-talker automatic speech recognition (ASR) that eliminates the need for explicit speaker queries. Unlike conventional approaches requiring target speaker embeddings or enrollment audio, our technique dynamically adapts individual ASR instances through speaker-wise speech activity prediction. The key innovation involves injecting speaker-specific kernels generated via speaker supervision activations into selected ASR encoder layers. This enables instantaneous speaker adaptation to target speakers while handling fully overlapped speech even in a streaming scenario.~Experiments show state-of-the-art performance in both offline and streaming scenarios, demonstrating that our self-adaptive method effectively addresses severe speech overlap through streamlined speaker-focused recognition. The results validate the proposed self-speaker adaptation approach as a robust solution for multi-talker ASR under severe overlapping speech conditions.
\end{abstract}

\section{Introduction}
Recent advancements in Automatic Speech Recognition (ASR), driven by improved architectures and larger training datasets, have significantly advanced the field. Concurrently, interest in multi-talker ASR has grown, particularly for applications such as analyzing natural conversations, developing voice assistants, and transcribing speech in health and legal contexts. Although this task is referred to by various terms—such as multi-talker ASR, multi-speaker ASR, or sometimes speaker-attributed ASR—the core challenge remains the same. Regardless of the terminology, transcribing speech signals in the presence of overlapping speech from multiple speakers is a demanding task, as ASR systems need to handle significantly increased variability.

Since the early days of ASR research, one of the most significant challenges for ASR systems has been modeling intrinsic and extrinsic variability in speech, often caused by speaker-specific factors such as accent, age, or gender. To tackle these challenges, speaker adaptation techniques were developed to address these variations. These techniques include the use of auxiliary speaker embeddings~\cite{saon2013speaker}, such as i-vectors~\cite{dehak2010front}, which represent speaker-specific traits and are integrated as additional features. Other methods involved feature transformation techniques, such as feature-space maximum likelihood linear regression (fMLLR)~\cite{abdel2013fast} and vocal tract length normalization (VTLN)~\cite{lee1996speaker}, which aim to generate speaker-independent features. Additionally, model-based adaptation techniques, such as learning hidden unit contributions (LHUC)~\cite{swietojanski2014learning} or linear transforms applied to neural network layers~\cite{zhang2015parameterised,zhao2016low}, were employed to capture speaker variability within the acoustic modeling framework. More recently, speaker adaptation techniques have been extended to end-to-end ASR models~\cite{baskar2022speaker,deng22_interspeech}. 
\begin{figure}[t]
  \centering
  \includegraphics[width=0.9\linewidth]{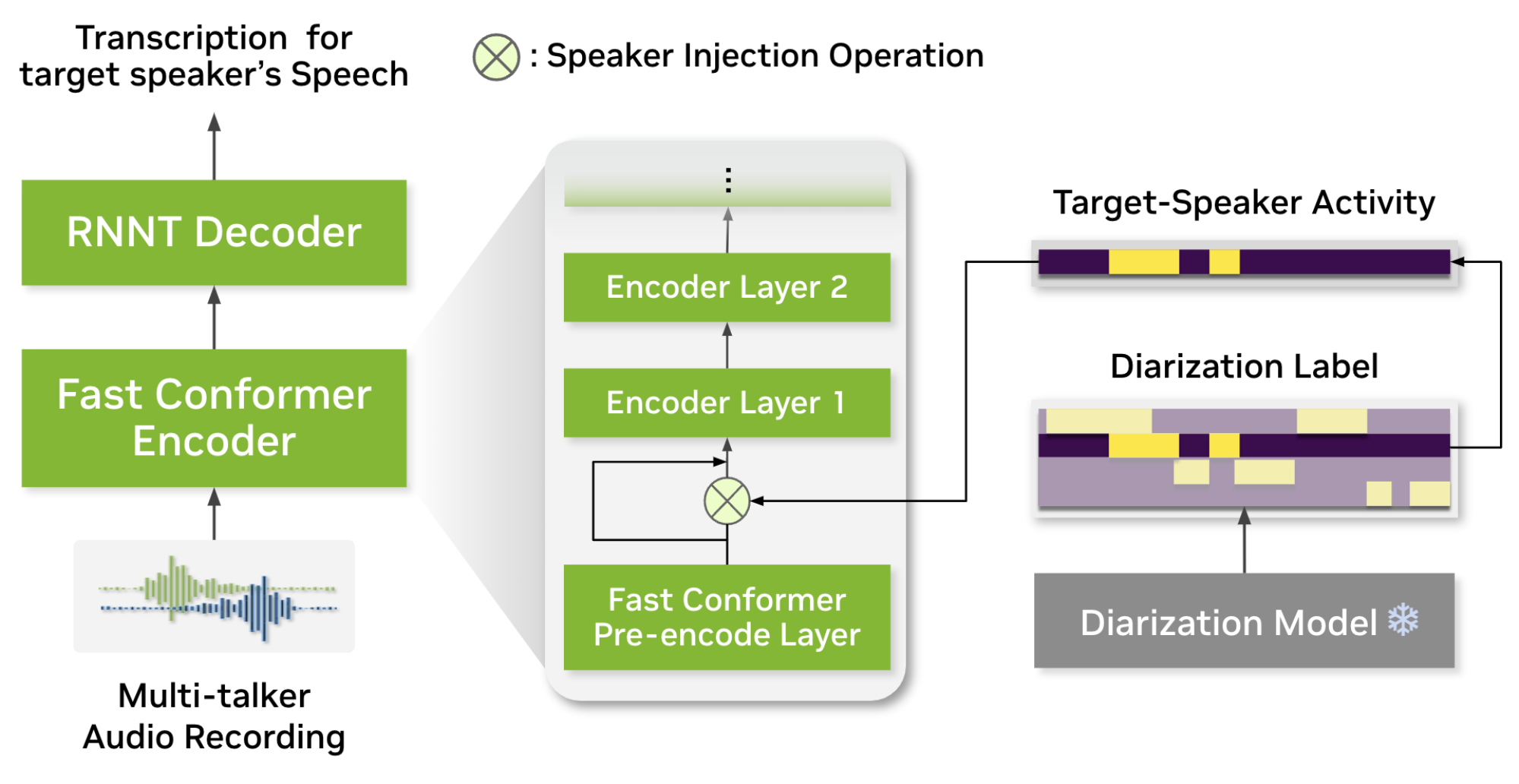}
  \vspace{-2ex}
  \caption{Speaker injection at pre-encode layer of Fast Conformer encoder.}
  \label{fig:speaker_injection}
  \vspace{-4ex}
\end{figure}

In multi-talker scenarios, handling frequent overlapped speech poses a significant challenge. To address this, speaker diarization systems are commonly employed to detect and separate the speech of individual speakers. The separated segments are then passed to a single-speaker ASR model for transcription \cite{medennikov2020stc}. Techniques such as guided source separation (GSS) are often used to estimate spectral masks~\cite{boeddeker2018front}, facilitating the separation process.~A major breakthrough in end-to-end multi-speaker ASR was achieved with the introduction of Serialized Output Training (SOT) in~\cite{kanda2020serialized}. Unlike earlier multi-talker ASR systems~\cite{chang2019end, chang2019mimo}, which relied on multiple encoders or heads, SOT serializes overlapped speech, enabling the multi-head self-attention mechanism~\cite{vaswani2017attention} to align speech signals with their corresponding tokens effectively. Subsequent advancements have led to numerous improved versions of the SOT approach~\cite{kanda2020joint,sotdom2024,fan2024sa}.

In parallel, speaker-attributed ASR~\cite{kanda2021end} has significantly improved ASR performance for overlapped multi-talker speech. Recently, alignment-free training (AFT)~\cite{moriya2024alignment} has shown that high performance can be achieved without explicit alignment by training and predicting each speaker's speech separately. These advancements mark substantial progress in multi-talker ASR. In addition, a more sophisticated but streamlined target-speaker-based approach, called Diarization-Conditioned Whisper (DiCoW)~\cite{polok2026dicow}, was proposed, in which diarization results condition the ASR model through a query-key biasing scheme.


In this paper, we propose a self-speaker adaptation (SSA) technique that can be effectively repurposed for multi-talker automatic speech recognition (ASR). A common challenge in single speaker streaming ASR systems is their tendency to prioritize a specific speaker—often the one closest to the microphone or the first to appear—while disregarding other speakers in multi-talker scenarios. This behavior arises from the encoder states being optimized to maximize accuracy for a particular speaker, which poses a significant challenge when fine-tuning a single-speaker ASR model for multi-talker applications. Specifically, the fine-tuned model must undergo substantial weight adjustments to counteract this inherent bias and achieve balanced recognition across all speakers. 

On the contrary, we leverage this mechanism in reverse to reinforce the propensity of adhering to a specific speaker. Our proposed technique achieves this by injecting a learnable speaker kernel into the pre-encode layer of the ASR encoder~\cite{rekesh2023fast}. This enables the ASR encoder to detect speech presence and utilize the speech kernel to adapt the encoder states dynamically. As a result, the encoder becomes more responsive to the targeted speaker's speech characteristics. However, this approach requires deploying one model instance per speaker, meaning the number of model instances must match the number of speakers. While this necessitates additional computational resources, it significantly enhances multi-talker ASR performance, achieving state-of-the-art error rates on benchmark datasets for multi-talker ASR.

Unlike traditional speaker adaptation methods that rely on external speaker representations, such as i-vectors~\cite{dehak2010front,saon2013speaker} or neural speaker embeddings~\cite{kanda2021end}, our technique depends solely on the speech activity of a specific speaker. For this reason, we refer to our method as self-speaker adaptation. 
In the experimental section, we compare our proposed method with baseline single-speaker ASR models and other studies reported on the same benchmark datasets, evaluating performance in both offline and streaming multi-talker ASR scenarios.


\section{Proposed Method}

Typically, target-speaker ASR systems rely on target speaker embeddings or enrollment audio to extract and utilize speaker-specific information from multi-talker utterances. However, the performance of such systems is highly dependent on the quality of the provided queries. For instance, a clean and noise-free query audio is generally preferred to achieve optimal results in target-speaker ASR tasks. In this work, we introduce a novel SSA approach that enables the model to adapt to a specific speaker using only the corresponding speech activities, eliminating the need for high-quality external queries. This method leverages the inherent speaker characteristics present in the input audio, allowing the model to dynamically adjust its focus to better recognize the target speaker's speech.~By doing so, our approach reduces the dependency on external resources and enhances the robustness of the system in real-world scenarios where high-quality queries or speaker representations may not always be available.

\subsection{Self-Speaker Adaptation}

As illustrated in Figure \ref{fig:speaker_injection}, our proposed method incorporates a speaker injection module into one of the layers of a single-speaker ASR model. Specifically, the speech activity is treated as a mask and applied to the output of the selected layer, with a residual connection added to preserve the original information. This process can be formally expressed as:
\begin{align}
    \mathbf{X}^i_\text{inj} = f_\text{inj}(\mathbf{X}^i, \mathbf{y}_{\text{spk}_k}) + \mathbf{X}^i,
\end{align}
where $\mathbf{X}^i \in \mathbb{R}^{B \times T \times D}$ is the i-th layer output, $\mathbf{y}_{\text{spk}_k} \in (0, 1)^{B \times T \times 1}$ is the corresponding speech activity for k-th speaker, and $\mathbf{X}^i_\text{inj}$ is the output with the speaker information injected. $B$, $T$ and $D$ stand for batch size, number of frames and feature dimension, respectively. Here, $f_\text{inj}$ can be any module that injects the speaker information (\textit{i.e.,} learnable speaker kernel) into the layer output. In this paper, two linear layers with an activation in between were applied for simplicity:
\begin{align}
    f_\text{inj}(\mathbf{X}^i, \mathbf{y}_{\text{spk}_k}) = f_\text{feedforward}(\mathbf{X}^i\odot\mathbf{y}_{\text{spk}_k}).
\end{align}
During the training stage, each input utterance contains overlapping speech from multiple speakers, and the speech activity of the target speaker is provided as an additional input. The target speaker is randomly selected from the set of speakers present in the utterance. Given the speech activity of the selected speaker, the model adapts to the target speaker autonomously, without requiring any pre-registered speaker profile. This design is particularly effective for streaming inference, where obtaining speaker profiles from overlapping speech is extremely difficult. The speech activities used during training can be obtained either from the ground truth labels or by applying a speaker diarization model to the input audio.

By leveraging the provided speech activities, the model dynamically adapts to the specific speaker, enabling robust recognition even in multi-speaker scenarios.~Notably, the training process remains nearly identical to that of a conventional single-speaker ASR system, including the training objective and decoding procedure. The only distinctions lie in the model architecture, which includes the speaker injection module, and the use of multi-speaker speech data for training. This approach ensures compatibility with existing ASR frameworks while enhancing the model's ability to handle target-speaker scenarios effectively.

\begin{figure}[t]
  \centering
  \includegraphics[width=\linewidth]{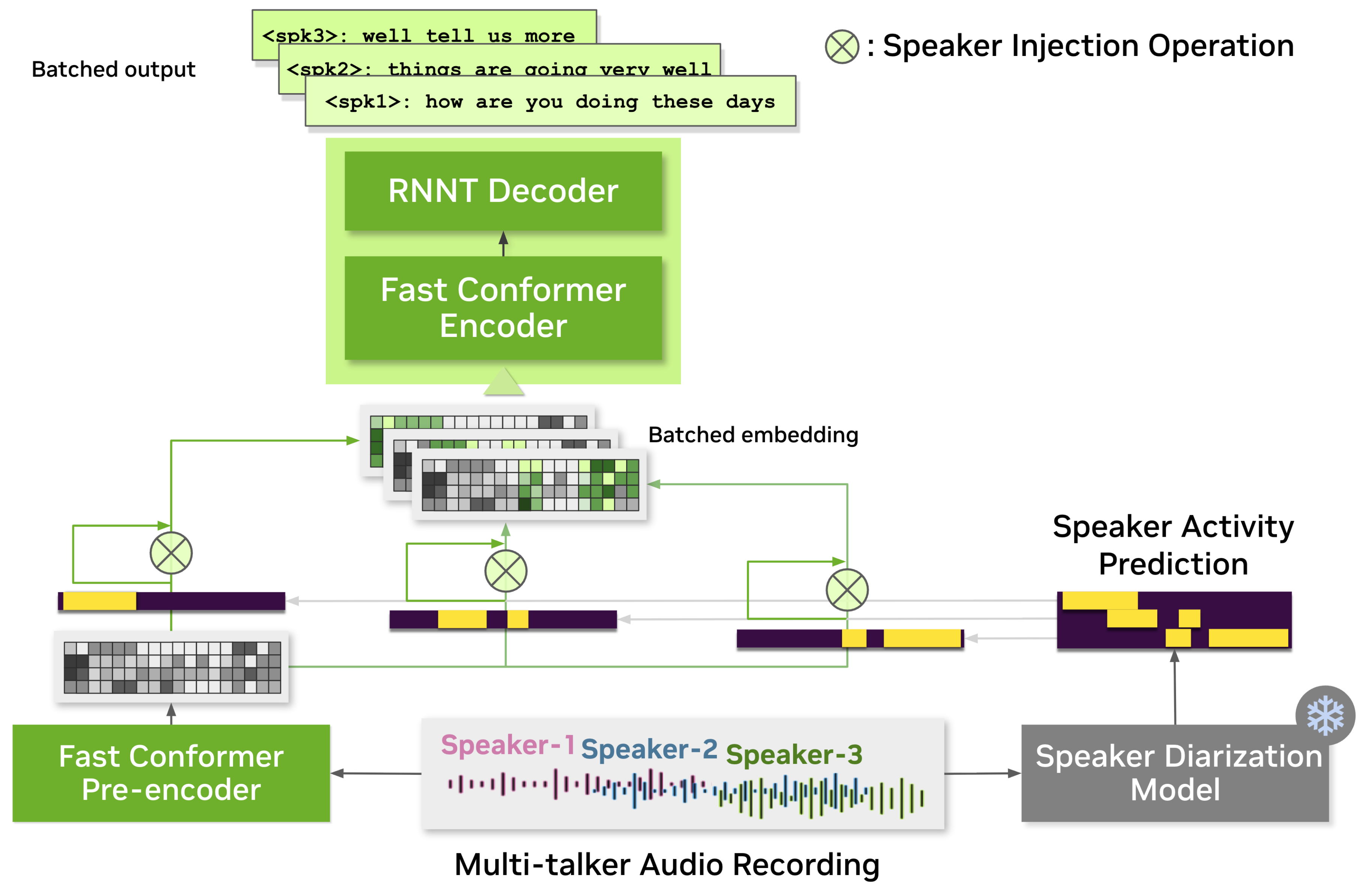}
  \vspace{-4ex}
  \caption{Multiple model instances are created and each model instance focuses on each speaker in the multi-talker recording.}
  \label{fig:multi_instance}
  \vspace{-4ex}
\end{figure}

\subsection{Repurposing for Multi-Talker ASR}


The proposed model is designed to adapt to a specific speaker when the corresponding speech activities are provided. This flexibility allows the model to serve dual purposes depending on the source of the speech activity information. If the speech activities are obtained from a personal VAD model \cite{ding20_odyssey}, the model functions as a target-speaker ASR system, operating in a way that is consistent with the training process. Alternatively, when a speaker diarization system is used to generate speaker-specific speech activities, the model can be extended to a multi-speaker ASR system by running multiple instances in parallel, each focusing on a different speaker identified by the diarization output, as illustrated in Figure \ref{fig:multi_instance}. Unlike conventional TS-ASR systems that require a speaker profile in advance, our model does not rely on any speaker identity information. Instead, it uses only the speech activity of the target speaker to adapt to that speaker during inference. This feature makes the model especially suitable for streaming applications, where acquiring a reliable speaker profile is often difficult or infeasible, particularly when the speaker begins with overlapped speech. As a result, the proposed method provides a practical and efficient solution for both single-speaker and multi-speaker ASR in real-time scenarios.

Although the model is trained similarly to a single-speaker ASR system, meaning it can only focus on one speaker at a time during each inference step, it can be adapted to handle multiple speakers when multiple speech activities are available. These activities are typically provided by a speaker diarization model. To decode speech from multiple speakers simultaneously, multiple instances of the model can be employed. Specifically, a batch processing approach can be used, where the batch size corresponds to the number of speakers in the input audio. Each instance within the batch processes the speech activity of a distinct speaker, enabling the model to generate transcriptions for all speakers concurrently. This approach maintains the simplicity of the single-speaker ASR framework while extending its functionality to multi-talker scenarios.

\begin{figure}[t]
  \centering
  \includegraphics[width=\linewidth]{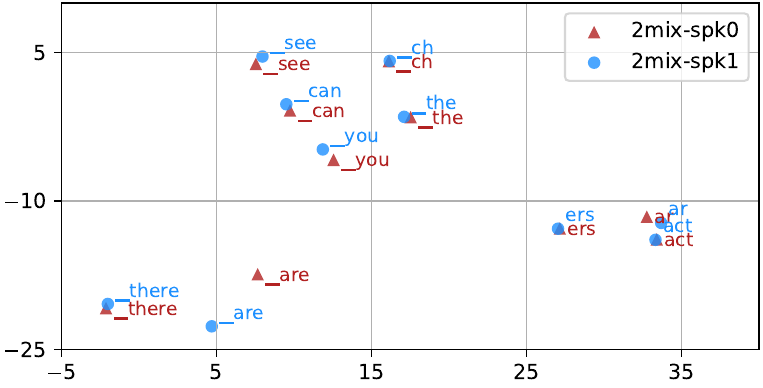}
  \vspace{-3ex}
\caption{t-SNE plot of Fast Conformer's last encoder states (embeddings) from an identical audio input (``there are the characters you can see''), using the same model weights but processed with different speaker kernels. The plot contrasts how ASR embeddings differ when generated using kernel $spk0$ (triangles) versus $spk1$ (circles), likely reflecting distinct speaker focuses.}
  \vspace{-2ex}
  \label{fig:tsne}
  \vspace{-2ex}
\end{figure}

Figure~\ref{fig:tsne} shows the t-SNE plot of the ASR encoder state (activation at the last layer) for each token, where \textit{2mix-spk0} and \textit{2mix-spk1} are injected with the first and second speaker's kernel, respectively. It is important to note that these ASR encoder representations, \textit{2mix-spk0} and \textit{2mix-spk1}, are derived from the same model and audio recording. We observe that the variability introduced by speaker injection is smaller than the distance between tokens; however, there is still a clear distinction between each speaker's kernel, enabling the model to decode overlapping speech from two or three speakers.

\subsection{Streaming Extension}
The proposed method can also be extended to streaming scenarios, provided that both the ASR model and the speaker diarization model support streaming capabilities. In this work, we employ the FastConformer Transducer model \cite{rekesh2023fast} with cache-aware streaming \cite{noroozi2024stateful} as the backbone for ASR training, ensuring efficient and low-latency processing of audio streams. In addition, we utilize the streaming Sortformer model \cite{streamingsortformer}, which is specifically designed for real-time end-to-end speaker diarization. By integrating these streaming-compatible models, the proposed method can be seamlessly adapted to streaming applications with minimal latency. This extension makes the system suitable for real-world applications such as live transcription, video conferencing, and other scenarios where low-latency processing is critical. The combination of streaming ASR and diarization models ensures that the system can handle continuous audio input while maintaining high accuracy.

\section{Experiments and Results}

\subsection{Datasets and Evaluation Metrics}
The training dataset was simulated using the LibriSpeech Corpus \cite{panayotov2015librispeech}. Since the model was trained with a single-speaker objective, alignment was not required for timestamp generation during training. For evaluation, we utilized the LibriSpeechMix dataset \cite{kanda2020serialized}, which includes 1-mix, 2-mix, and 3-mix data to simulate single-speaker, two-speaker, and three-speaker scenarios, respectively. We also finetune the model on Fisher English Training Speech Part 1 and 2~\cite{cieri2004fisher} and evaluate it on CH109, a two-speaker subset of 109 sessions from the Callhome American English Speech~(CHAES) dataset~\cite{canavan1997CALLHOME}.

To evaluate the performance of the proposed model, we employed the concatenated minimum-permutation word error rate (cpWER) metric \cite{watanabe2020chime}, which is commonly used for multi-talker ASR systems. This metric ensures a fair comparison by considering the best possible alignment between the predicted and reference transcriptions across all speakers.

\begin{table}[t!]
\centering
\vspace{-1ex}
\caption{DER results on LibriSpeechMix and CH109.}
\vspace{-2ex}
\label{table:der}
\setlength{\tabcolsep}{5pt} 
\begin{tabular}{r|c|c|ccc}
\Xhline{2\arrayrulewidth}
\toprule
 &  \multirow{2}{*}{Size} & Latency & \multicolumn{3}{c}{DER (\%)} \\
 & & (ms) & 2-mix & 3-mix & CH109 \\
\midrule
 \multirow{2}{*}{Sortformer} & \multirow{2}{*}{123M} &  $\infty$  & 2.5 & 3.2 & - \\ 
 & & 1120  &  3.3 & 4.1 & 5.6 \\ 
 \bottomrule
\Xhline{2\arrayrulewidth}
\end{tabular}
\vspace{-1ex}
\end{table}

\subsection{Training Details}
\label{sec:train}
For the ASR model, we employed the FastConformer Transducer model \cite{rekesh2023fast,noroozi2024stateful} and initialized it using the publicly available pre-trained model
\footnote{Offline ASR model: \url{huggingface.co/nvidia/stt_en_fastconformer_transducer_large}}\footnote{Streaming ASR model: \url{huggingface.co/nvidia/stt_en_fastconformer_hybrid_large_streaming_multi}}. For the streaming diarization model, we use a fine-tuned version of the streaming Sortformer model\footnote{Streaming diarization model: \url{huggingface.co/nvidia/diar_streaming_sortformer_4spk-v2}} \cite{park2024sortformer} on the simulated training set derived from the LibriSpeech Corpus. 

The performance of the diarization model is reported in Table \ref{table:der}, which shows the Diarization Error Rate (DER) for LibriSpeechMix 2-mix, 3-mix with a collar of 0, and CH109 datasets with a collar of 0.25. For LibriSpeechMix evaluation, the silence at the beginning and end of each utterance is ignored. 

During training, the diarization model was kept frozen and used to provide speech activities for each speaker. The speech activities were randomly sampled from one of the speakers in the input audio, and the corresponding transcriptions for that speaker were used as the training labels. This approach ensures that the model learns to adapt to the target speaker dynamically without requiring additional speaker-specific information. The top five checkpoints with the lowest validation word error rate (WER) are selected and averaged to produce the final checkpoint used for evaluation.

Both streaming and offline models were trained for 200k steps. For tokenization, we used SentencePiece \cite{kudo2018sentencepiece} with byte pair encoding (BPE) and a vocabulary size of 1024. The tokenizer was trained on the training set of each dataset to ensure compatibility with the input data. The models were optimized using the AdamW optimizer \cite{loshchilov2017decoupled} with a weight decay of 0.001 and the Noam learning rate scheduler \cite{vaswani2017attention} with a coefficient of 5.0. All training runs were conducted with 8×NVIDIA Tesla A100 GPUs on one node. 

\subsection{Finetuning details}
\label{sec:finetune}
After training on the LibriSpeechMix dataset, we further fine-tune the model on the Fisher dataset for an additional 50k steps. Each session is truncated into segments ranging from 10 to 30 seconds, resulting in 64k training utterances and 1k validation utterances. The top five checkpoints with the lowest validation WER are selected and averaged to produce the final checkpoint used for evaluation.

\subsection{Evaluation and Comparative Analysis}

\subsubsection{Evaluation on LibriSpeechMix}
Offline model is trained with 1-mix, 2-mix, and 3-mix simulated data, with a ratio of 1:3:6. We report the evaluations on 1-, 2-, and 3-mix data in Table~\ref{table:offline_max3}. The proposed SSA model achieves state-of-the-art performance on the LibriSpeechMix dataset, with cpWERs of 2.2\%, 2.8\%, and 5.0\% for 1-mix, 2-mix, and 3-mix scenarios, respectively. It outperforms most of the strong baselines with up to 3 speakers in the training data, particularly excelling in complex multi-talker settings. The model contains 238M parameters, which consist of a FastConformer Transducer ASR model (114M), a Sortformer diarization model (123M) and a speaker injection module (1.1M). The results highlight the effectiveness of the SSA mechanism in handling overlapping speech, making SSA a highly competitive and scalable solution for multi-talker ASR tasks.

\begin{table}[t!]
\centering
\vspace{-2ex}
\caption{Offline Max.~3-speaker systems on LibriSpeechMix.}
\vspace{-2ex}
\label{table:offline_max3}
\setlength{\tabcolsep}{3pt} 
\begin{tabular}{rc|ccc}
\Xhline{2\arrayrulewidth}
\toprule
        & Model &\multicolumn{3}{c}{cpWER (\%)}   \\
Systems & Size  & 1-mix & 2-mix & 3-mix    \\
\midrule
SOT~\cite{kanda2020serialized}           &  135.6M & 4.6  & 11.2 & 24.0 \\
SOT-SQR~\cite{kanda2020joint}            &  136M   & 4.2  & 8.7  & 20.2 \\ 
E2E-SA~\cite{kanda2021end}               &  128.6M & 3.3  & 4.3  & 6.0  \\
Sidecar-Sep~\cite{meng2023sidecar}       &  103.6M & -    & 5.7  & -   \\ 
MT-Whisper-L ~\cite{meng24c_interspeech} &  1.56B  & -    & 3.4  & 6.8  \\
DOM-SOT~\cite{sotdom2024}                &  33M    & 5.2  & 5.6  & 10.0 \\
SA-SOT~\cite{fan2024sa}                  &  136M   & 3.4  & 8.2  & -   \\ 
MT-LLM~\cite{mtllm2024}                  &  8B     & 2.3  & 5.2  & 10.2 \\
AFT-MT~\cite{moriya2024alignment}        &  156M   & 2.4  & 3.4  & -   \\ 
\midrule
(Proposed) \textbf{SSA}              & 238M & \textbf{2.2} & \textbf{2.8} & \textbf{5.0}  \\ 

\bottomrule
\Xhline{2\arrayrulewidth}
\end{tabular}
\vspace{-3ex}
\end{table}

For the streaming model, since there are no existing results for the 3-mix LibriSpeechMix dataset for comparison, we trained two distinct models. The first model is trained using 1-mix and 2-mix simulated data, with a ratio of 1:9 between 1-mix and 2-mix samples. The second model is trained using 1-mix, 2-mix, and 3-mix simulated data, with a ratio of 1:3:6. Table~\ref{table:streaming_max2} presents the results of the model trained with up to 2 speakers, while Table~\ref{table:streaming_max3} shows the results of the model trained with up to 3 speakers. The proposed SSA scheme demonstrates strong performance in streaming multi-talker ASR, with cpWERs of 4.0\% (1-mix) and 5.6\% (2-mix) at 560 ms latency, outperforming baselines under highly overlapped scenarios. Table~\ref{table:streaming_max3} shows streaming multi-talker ASR with a maximum of 3 speakers with minor performance degradation from the 2-speaker model.

\begin{table}[t!]
\centering
\vspace{-2ex}
\caption{Streaming Max.~2-speaker systems on LibriSpeechMix.}
\vspace{-2ex}
\label{table:streaming_max2}
\setlength{\tabcolsep}{5.0pt} 
\begin{tabular}{r|c|cc|cc}
\Xhline{2\arrayrulewidth}
\toprule
      &   & \multicolumn{2}{c|}{Latency(ms)} & \multicolumn{2}{c}{cpWER (\%)} \\     
                                              & Size & ASR & Diar. & 1mix & 2mix \\
                                                           
\Xhline{2\arrayrulewidth}
\midrule
\multirow{2}{*}{Stream-T-SOT~\cite{kanda22b_interspeech}}  & \multirow{2}{*}{160M} & 160  & 2720 & 4.9 & 6.5  \\ 
                                                           & & 2560 & 2720 & 3.3 & 4.7  \\
SSL-BLM-MT~\cite{huang2023self}                            & - & 160  & -    & 7.2 & 9.6  \\
T-SOT-FNT~\cite{kanda2021end}                              & - & 160  & -    & 4.7 & 10.1 \\
AFT-MT~\cite{moriya2024alignment}                          & 156M & 640  & -    & 4.0 & 6.3 \\ 
\midrule
\multirow{5}{*}{(Proposed) \textbf{SSA}}    & \multirow{5}{*}{238M}     
    & \multicolumn{2}{c|}{80}   & 6.8      & 8.3 \\ 
    & & \multicolumn{2}{c|}{160}  & 5.7      & 7.2 \\ 
    & & \multicolumn{2}{c|}{560}  & \textbf{4.0}      & \textbf{5.6} \\ 
    & & \multicolumn{2}{c|}{1120} & 3.8 & 5.2 \\ 
    & & \multicolumn{2}{c|}{2720} & 3.4 & 4.6 \\ 

\bottomrule
\Xhline{2\arrayrulewidth}
\end{tabular}
\vspace{-1ex}
\end{table}

\begin{table}[htbp!]
\centering
\caption{Streaming Max.~3-speaker systems on LibriSpeechMix.}
\vspace{-1ex}
\label{table:streaming_max3}
\setlength{\tabcolsep}{5pt} 
\begin{tabular}{r|c|c|ccc}
\Xhline{2\arrayrulewidth}
\toprule
 &   & Latency & \multicolumn{3}{c}{cpWER (\%)} \\
  & Size  & (ms) & 1mix  & 2mix & 3mix   \\
\midrule
\multirow{5}{*}{(Proposed) \textbf{SSA}} & \multirow{5}{*}{238M} &  80  & 7.1 & 9.2 & 15.6 \\ 
 & & 160  & 6.2  & 7.8 & 15.7 \\ 
 & & 560  & 4.3  & 5.9 & 11.0  \\ 
 & & 1120  & 4.0  & 5.4 & 9.8  \\ 
 & & 2720  & 3.7  & 4.9 & 8.1  \\  
\bottomrule
\Xhline{2\arrayrulewidth}
\end{tabular}
\vspace{-1ex}
\end{table}


\subsubsection{Evaluation on Real data}
We also evaluate the proposed model on a real-life multi-speaker dataset CH109. The baseline is a cascade streaming multi-talker ASR model, which consists of a streaming single speaker ASR model and a streaming diarization model. In the cascaded model, words and speaker segments are matched using time-stamps. The ASR model for the cascaded baseline is the same as the model whose parameters are employed for initialization for streaming model training, as mentioned in Section \ref{sec:train}, and the streaming diarization model for the baseline is the same as the model used for streaming SSA inference, as mentioned in Section \ref{sec:finetune}.~The proposed SSA model demonstrates superior performance on the CH109 dataset, achieving a cpWER of 26.21\% at a latency of 1120 ms. This outperforms not only the cascaded streaming model but also the performance of the offline system \cite{park2024enhancing}, despite our proposed system operating under low-latency streaming constraints. 


\begin{table}[t!]
\centering
\caption{Streaming Max.~2-speaker systems on CH109.}
\vspace{-2ex}
\label{table:streaming_ch109}
\setlength{\tabcolsep}{5pt} 
\begin{tabular}{r|c|c|c}
\Xhline{2\arrayrulewidth}
\toprule
 &  Size & Latency (ms) & cpWER (\%) \\
\midrule
LLM-BSD \cite{park2024enhancing} & 2.2B & $\infty$  &  26.31 \\
 Cascaded Model & 238M & 1120  &  27.38\\
 \midrule
  (Proposed) \textbf{SSA} & 238M & 1120 & 26.21 \\ 
 \bottomrule
\Xhline{2\arrayrulewidth}
\end{tabular}
\vspace{-4ex}
\end{table}



\section{Conclusion}
In this paper, we introduced a query-less speaker targeting approach that employs a multi-instance encoder-decoder for each speaker. Following the design principle of maximizing the performance of the base monaural ASR system, the proposed multi-instance speaker targeting approach shows that the relatively long frame length can be addressed by employing multiple instances of single-speaker ASR models. Our method achieves state-of-the-art performance in both streaming and offline setups on the LibriSpeechMix dataset, effectively addressing the challenges of multi-speaker scenarios and overlapping speech. Future work includes the application of the proposed method on Transformer-based ASR systems and multi-modal large language models (LLMs), thus equipping the foundational ASR models or LLMs with the state-of-the-art multi-talker ASR capability.


\bibliographystyle{IEEEtran}
\bibliography{reference}

\end{document}